\documentclass{ptptex}

\usepackage{graphicx}
\usepackage{psfrag}

\def\nn{\nonumber}
\title{Characteristic Behavior of Toroidal Carbon Nanotubes}
\subtitle{Kinematics of Persistent Currents} 

\author{Ken-ichi Sasaki and Yoshiyuki Kawazoe}
\inst{Institute for Materials Research, Tohoku University, 
Sendai 980-8577, Japan}

\recdate{February 26, 2004}

\abst{
 The electrical properties of a carbon nanotube depend strongly on its
 lattice structure as defined by chiral and translational vectors.
 A toroidal shape for a nanotube allows various twisted structures to
 exist along the direction of the tube axis.
 We theoretically investigate the kinematics of conducting electrons and
 persistent currents in toroidal carbon nanotubes.
 We show that persistent currents exhibit a special characteristic of
 the cylindrical lattice structure in twisted cases.
 We discuss the possibilities that the twist alters the period of the
 current to one half the flux quantum and that the current flows without
 an external magnetic field.
}

\notypesetlogo
\begin{document}

\maketitle
%%%%%%%%%%%%%%%%%%%%%%

\section{Introduction}\label{sec:introduction}

Carbon nanotubes~\cite{Iijima} (CNTs) are cylindrical molecules with
diameters as small as one nanometer and lengths up to several
micrometers.
They consist of carbon atoms only and can be thought of as graphene
sheets wrapped into cylinders. 
They exhibit either metallic or semiconducting behavior, depending on
the diameter and chirality of the hexagonal carbon lattice of the 
tube~\cite{SDD,MDW,Wildoer}.
It is quite important to understand the electrical properties of carbon
nanotubes, which are governed mainly by low-energy electrons near the
Fermi level.
Thus far, theoretical research has concentrated mainly on the low-energy
electrical properties of metallic nanotubes.
Because a nanotube is very long and thin, theorists have analyzed its
properties by regarding it as a one-dimensional object.
Such analyses, based on the method developed by Tomonaga and Luttinger
(TL-liquid theory), have revealed the nature of the correlated
system~\cite{Kane}.
The characteristic feature of the correlation was observed as a
power-law dependence of the resistivity on temperature~\cite{Bockrath}.
Taking into account the successful application of TL-liquid theory to
CNTs, in terms of their low-energy electrical behavior, CNT systems seem 
very similar to conventional one-dimensional materials such as a chain
of atoms.
The following question then naturally arises:
How does a nanotube differ from conventional one-dimensional materials,
such as a chain of atoms?
Concerning this point, we note the persistent currents in toroidal
carbon nanotubes~\cite{Liu}.
A tube curved so that both ends connect in a toroidal form is hereafter
referred to as a ``torus'' or ``nanotorus'' for simplicity.

Persistent currents in mesoscopic rings are known phenomena in
condensed matter physics~\cite{BIL,Webb,Imry}.
Conservation of the electron phase coherence in an entire sample can 
affect the equilibrium properties of the system. 
One of the most striking consequences of this is that a single
isolated mesoscopic normal-metal ring threaded by a magnetic flux is
thought to carry a (persistent) current in the form of a sawtooth curve
with period $\Phi_0$, the single-electron flux quantum. 
In this paper, we show that nanotori can exhibit special persistent
currents not seen in usual mesoscopic systems.
For example, some types of nanotori exhibit a sawtooth curve with a
period one half the flux quantum. 
We point this out as an effect reflecting the geometrical structure of
the (graphite) cylinder; that is, conducting electrons have a new degree
of freedom to rotate around the tubule axis that is not found in
conventional one-dimensional materials such as a chain of atoms.

Here, we would like to mention the relationship between our work and
previously published literature on persistent currents in carbon
nanotori.
Lin and Chuu~\cite{LC} carried out numerical estimations of persistent 
currents in untwisted nanotori and found structure dependent currents.
Subsequently, the non-trivial geometrical degree of freedom constituted
by {\it twist}~\cite{Ceulemans} was taken into account by Marga\'nska
and Szopa~\cite{LC}. 
From a numerical calculation, they concluded that for a specific type
of twisted torus, the twist has no significant influence on the
persistent current. 
The above-mentioned authors employed a simple nearest-neighbor
tight-binding Hamiltonian that was shown to be a good approximation for
describing the conducting electrons in nanotubes~\cite{SDD}.
In this paper, we use the same Hamiltonian and examine persistent
currents in all possible types of torus geometries analytically and
consider the possibility that carbon nanotori exhibit unusual (and new) 
phenomena.
We clarify two interesting possibilities:
(1) The twist can change the period of the current to one half the flux
quantum, and (2) current may flow without an external magnetic field.

One can include various effects on persistent currents in the analysis. 
The curvature effect was analyzed by Lin et al.~\cite{LRR}, and 
the effect of disorder on persistent currents was taken into account by
Latil et al.~\cite{LRR} as a position-dependent on-site energy. 
The effect of Coulomb interactions was examined by Odintsov et
al.~\cite{LRR} and Sasaki~\cite{LRR}.
Liu et al.~\cite{Liu} observed toroidal structures experimentally.
Martel et al. and Ahlskog et al.~\cite{Martel} suggested that they are
likely to be coiled nanotubes stabilized by van der Waals interactions.

This paper is organized as follows.
In \S\S~\ref{sec:normal_torus} and \ref{sec:PCinNT}, we review the
basis for the kinematics of conducting electrons in untwisted nanotori
and the persistent currents in those systems. 
We examine persistent currents not only in metallic structures but also
in semiconducting chiral structures, assuming that a finite number of
states exist near the Fermi level. 
(While this is true for metallic nanotori, it is not true for
half-filled semiconducting nanotori.
However, given that a sufficient number of electrons are added to the
system, it is possible to observe the persistent current.)
In \S\S~\ref{sec:twisted_torus} and \ref{sec:PCinTT}, we clarify the
kinematics of electrons in twisted nanotori and examine persistent
currents.
We study the effects of the twist on persistent currents and show that,
due to the cylindrical lattice structure of the nanotube, a special
current can flow in these systems.
In \S~\ref{sec:discussion} we summarize and discuss our results,
and in \S~\ref{sec:conclusion} we give our conclusions.
For the purposes of this study, we use in units for which $\hbar = c =
1$.

\section{Kinematics of an untwisted torus}\label{sec:normal_torus}

We begin by specifying the lattice structure of a nanotorus.
A nanotorus is a nanotube whose ends are connected.
A nanotube is a graphene sheet wrapped to form a cylinder. 
Thus, a nanotorus can be classified according its chiral and
translational vectors defined, respectively, by
\begin{equation}
 C_h = n T_1 + m T_2, \ \ 
  T = p T_1 + q T_2, 
  \label{eq:chiral-vector}
\end{equation}
where $T_1$ and $T_2$ are the symmetry translation vectors on the planar 
honeycomb lattice.~\footnote{A schematic diagram of the honeycomb lattice 
and the notation used in this paper can be found in Ref.~\citen{LRR}
(Sasaki). 
The symmetry translation vectors can be expressed as 
$T_1 = \sqrt{3} ae_x,
T_2 = (\sqrt{3}/2) ae_x + (3/2) a e_y$, 
where $a$ denotes
the distance between two nearest carbon sites, and each site can be
reached from any other site through a translation consisting of
combination of the vectors $u_a \ (a = 1,2,3)$.
These vectors are given explicitly by 
$ u_1 = a e_y,  
  u_2 = -(\sqrt{3}/2)a e_x - (1/2) a e_y,
  u_3 = (\sqrt{3}/2)a e_x - (1/2) a e_y.
$}
The two sets of integers $(n,m)$ and $(p,q)$ specify the lattice
structure around and along the axis, respectively.
Note that, in the case of a torus, the chiral vector does not determine
the translational vector, in contrast with the case of nanotubes, where 
the integers $(n,m)$ completely fix the unit of the translational
vector: $(p,q)/{\gcd}(p,q)$~\cite{SDD}.
Here, ${\rm gcd}(p,q)$ represents the greatest common divisor of the two 
integers $p$ and $q$.

\begin{figure}[htbp]
 \begin{center}
  \psfrag{a}{(a)}
  \psfrag{b}{(b)}
  \psfrag{C_h}{$C_h$}
  \psfrag{T}{$T$}
  \psfrag{A_C}{$A_C$}
  \psfrag{A_T}{$A_T$}
  \psfrag{A}{$A$}
  \psfrag{u}{$u$}
  \psfrag{d}{$d$}
  \includegraphics[scale=0.5]{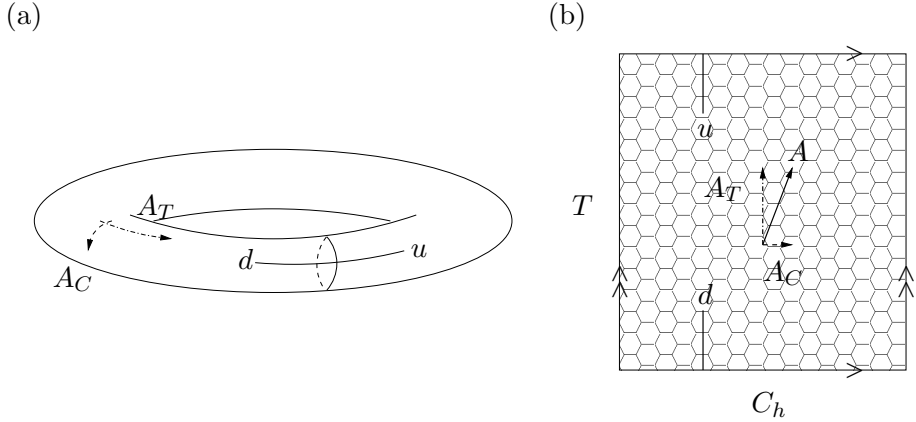}
 \end{center}
 \caption{Schematic diagram of an untwisted torus (a) and its net (b)
 with an external gauge field.
 The two lines extending upward from `$u$' and downward from `$d$' join
 to form an untwisted torus.}  
 \label{fig:normal_torus}
\end{figure}

For an {\it untwisted} torus we define chiral and translational vectors
that satisfy
\begin{equation}
 C_h \cdot T = 0.
  \label{eq:normal-torus}
\end{equation}
This condition ensures that there is no {\it twist} along the axis, so
that $(p,q)/{\rm gcd}(p,q)$ is determined by the chiral vector only.
An untwisted torus can be unrolled into a rectangular graphene sheet, as
is shown in Fig.~\ref{fig:normal_torus}.
We classify nanotori for which the translational vector does not satisfy 
Eq.~(\ref{eq:normal-torus}) as a {\it twisted} torus, which we will
investigate in later sections of this paper. 
Making use of Eqs.~(\ref{eq:chiral-vector}) and (\ref{eq:normal-torus}),
we can rewrite $(p,q)$ for an untwisted torus as
\begin{equation}
 \frac{p}{d_T} = \frac{2m+n}{d_R}, \ \
  \frac{q}{d_T} = -\frac{2n+m}{d_R},
\end{equation}
where $d_R \equiv {\rm gcd}(2m+n,2n+m)$ and $d_T \equiv {\rm gcd}(p,q)$.
By introducing the quantity $d \equiv {\rm gcd}(n,m)$, we obtain $d_R =
{\rm gcd}(3d ,n-m)$, a useful identity for classifying the lattice
structures of nanotubes~\cite{SDD} and also for understanding persistent
currents in nanotori.
%%%%%%%%%%%%%%%%%%%%%%%%%%%%%
%%%%%%%%%%%%%%%%%%%%%%%%%%%%%

The quantum mechanical states of the conducting electrons
($\pi$-electrons) are labeled by the wave vectors. 
The wave vector is fixed by the lattice structure of the nanotorus
through the boundary conditions. 
We define unit wave vectors $k_\perp$ and $k_\parallel$ for an untwisted
torus by the periodic boundary conditions
\begin{equation}
 C_h \cdot k_\perp = 2\pi,\ C_h \cdot k_\parallel = 0,\ 
  T \cdot k_\perp = 0,\ T \cdot k_\parallel = 2\pi.
\end{equation}
Here, $k_\perp$ and $k_\parallel$ are expressed in terms of the
reciprocal lattice vectors ($K_1$ and $K_2$) as
\begin{equation}
 k_\perp = \frac{1}{N_c} (-q K_1 + p K_2), \
  k_\parallel = \frac{1}{N_c} (m K_1 - n K_2),
  \label{eq:k_parallel_normal}
\end{equation}
where $N_c \equiv mp-nq$, and we have defined the reciprocal lattice
vectors of the graphene sheet by $K_i \cdot T_j = 2\pi \delta_{ij} \
(i,j = 1,2)$. 
It should be noted that $|N_c|$ corresponds to the total number of
hexagons in an untwisted nanotorus, as seen from the relation 
$T \times C_h = N_c (T_1 \times T_2)$, and it is an even number. 
We decompose the wave vector $k$ as 
$\mu_\perp k_\perp + \mu_\parallel k_\parallel$,
where $\mu_\perp$ and $\mu_\parallel$ are integers lying within the
Brillouin zone:
\begin{equation}
 \left[ -\frac{d}{2} \right]+1 \le \mu_\perp \le \left[ \frac{d}{2} \right], \ 
  \left[ -\frac{N_c}{2d} \right]+1 \le \mu_\parallel \le \left[ \frac{N_c}{2d} \right],
  \label{eq:Brillouin}
\end{equation}
where $[n]$ represents the largest integer smaller than $n$.
It should also be noted that wave vectors that are {\it congruent}, can
be identified with the same state.
Mathematically, two wave vectors $k$ and $k+\delta k$ are congruent if
$\delta k$ can be written as
\begin{equation}
 \delta k \equiv \delta \mu_\perp k_\perp + \delta \mu_\parallel k_\parallel
  = \tau_1 K_1 + \tau_2 K_2,
  \label{eq:cong-tau}
\end{equation}
where $\tau_1$ and $\tau_2$ are integers.
It follows from Eqs.~(\ref{eq:k_parallel_normal}) and
(\ref{eq:cong-tau}) that 
\begin{equation}
 \delta \mu_\perp = n \tau_1 + m \tau_2, \
  \delta \mu_\parallel = p \tau_1 + q \tau_2.
\end{equation}
Using this degree of freedom, the first inequality in
Eq.~(\ref{eq:Brillouin}) is equivalent to $1 \le \mu_\perp \le d$.
Nevertheless, we use Eq.~(\ref{eq:Brillouin}), because the wave vector
$\mu_\perp k_\perp$ represents the momentum around the axis, which
should take both positive and negative (or zero) values.
The orbital motion of an electron about the axis is represented by
a negative value of $\mu_\perp$ for clockwise motion and a positive 
$\mu_\perp$ for counterclockwise motion, which is manifest in
Eq.~(\ref{eq:Brillouin}).

\section{Persistent currents in an untwisted torus}\label{sec:PCinNT}

In this section, we consider persistent currents in untwisted nanotori.
We assume that the Hamiltonian for the $\pi$-electrons in an external
gauge field $A$ is the nearest-neighbor tight-binding Hamiltonian,
\begin{equation}
 {\cal H} =  V_\pi \sum_{\langle i,j \rangle} 
  a_j^\dagger e^{-ie\int_{r_i}^{r_j} A \cdot ds} a_i,
  \label{eq:hamiltonian}
\end{equation}
where $V_\pi$ is the hopping integral, and the sum is over pairs of
nearest-neighbor carbon sites $i,j$ on the surface. 
The vector $r_i$ is that pointing toward the site $i$, and $a_i$
and $a_i^\dagger$ are the canonical annihilation-creation operators of
the site $i$ electron, which satisfy the anti-commutation relation 
$\{ a_i,a_j^\dagger \} = \delta_{ij}$. 
Finally, $-e$ is the electron charge and $ds$ is the differential line element on
the surface.
Generally, the gauge field has two components:
$A \cdot T = A_T$ and $A \cdot C_h = A_C$.
The quantity $A_T$ is the Aharonov-Bohm flux $\Phi$ penetrating the
ring, and $A_C$ corresponds to the magnetic flux within the surface of a
nanotorus (see Fig.~\ref{fig:normal_torus}).

\begin{figure}[htbp]
 \begin{center}
  \psfrag{+}{$+\mu_\perp^0 k_\perp$}
  \psfrag{-}{$-\mu_\perp^0 k_\perp$}
  \includegraphics[scale=0.45]{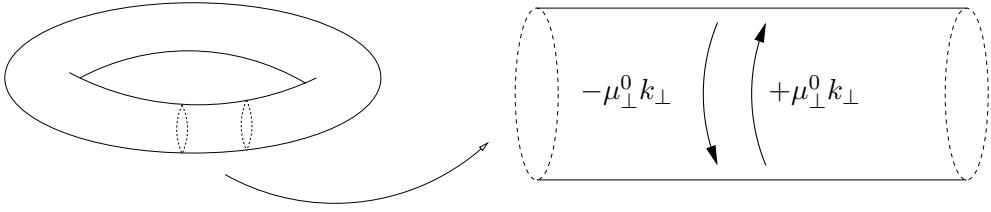}
 \end{center}
 \caption{A part of a nanotorus in which the motion of the electrons
 near the Fermi level are shown (axial motion is ignored).
 The arrows indicate two modes near the Fermi level, or the motion of 
 the electrons in the low energy bands.
 If $\mu_\perp^0$ is congruent to zero, there is no orbital motion about
 the axis.
 In this case, the wave function of the electrons can be regarded as a
 constant around the axis.}
 \label{fig:low_energy_modes}
\end{figure}

We diagonalize the Hamiltonian using the Bloch base states and obtain
the energy eigenvalue
\begin{equation}
 E(k-eA) = - V_\pi \left| \sum_{a=1,2,3} e^{i (k-eA) \cdot u_a} \right|,
\end{equation}
where $E(k-eA)$ is the energy eigenvalue below the Fermi level
($E \le 0$).
Here, the vectors $u_a \ (a = 1,2,3)$ form a triad pointing in the three
directions of the nearest neighbors of a carbon site.
For non-interacting theories, the persistent current can be calculated
form the behavior of electrons near the Fermi level~\cite{Imry}.
It is therefore convenient to select energy bands for which the
electronic states are located closest to the Fermi level.
Hereafter, we call these energy bands {\it low energy bands}.
By studying the energy dispersion relation of the Hamiltonian, we find
that there are two independent Fermi points. 
They are located at 
\begin{equation}
 \pm K + \tau_1 K_1 + \tau_2 K_2,
\end{equation}
where $K \equiv (2K_1 + K_2)/3$ satisfies $E(\pm K) = 0$, and
$(\tau_1,\tau_2 )$ is a pair of integers representing the congruent
degree of freedom.
It is easy to find the index $\mu_\perp$ of the low energy bands.
We denote them as $\pm \mu_\perp^0$, which are given by
\begin{equation}
 \mu_\perp^0 = \left\langle \frac{2n+m}{3} \right\rangle + n \tau_1 + m \tau_2.
\end{equation}
Here, $\langle a \rangle$ represents the integer closest to the value
$a$. 
In the following treatment, it is important to verify that the electrons
in the low energy bands are {\it orbiting} about the axis.
In other words, it is crucial to determine if the wave functions of
electrons in the low energy bands are constant around the axis (see
Fig.~\ref{fig:low_energy_modes}).
One may regard the electrons in the low energy band as {\it
non-orbiting} if $\mu_\perp^0$ is congruent to zero through an
appropriate choice of $\tau_1$ and $\tau_2$. 
The condition of a non-orbiting mode is therefore
\begin{equation}
 \exists (\tau_1,\tau_2) \in Z \ \cdot \ \mu_\perp^0 = 0.
  \label{eq:rotation}
\end{equation}
%%%%%%%%%%%%%%%%%%%%%%%%%%%%%
%%%%%%%%%%%%%%%%%%%%%%%%%%%%%

In Table~\ref{table:list}, we list several types of chiral structures
and indicate whether the electrons in the low energy bands are orbiting.
\begin{table}
 \caption{Untwisted nanotori and electron motion in the low energy bands.}
 \begin{center}
  \def\arraystretch{1.5}
  \begin{tabular}{|c|c|c|} \hline
   {\bf Chirality} & {\bf Index of Low Energy Bands} & {\bf ``Orbiting?''} \\ \hline
   $C_h = (n,m)$ & $\mu_\perp^0 = \langle \frac{2n+m}{3} \rangle + n \tau_1 + m \tau_2 $ & $\mu_\perp^0 \not\cong 0$ ? \\ \hline
   $(n,0)$ zigzag & $ \langle \frac{2n}{3} \rangle +n \tau_1 $ & Yes if $(n \ne 1)$ \\ \hline
   $(n,n)$ armchair & $n + n \tau_1 + n \tau_2 $ & No \\ \hline
   $(n,2n)$ chiral & $ \langle \frac{4n}{3} \rangle +n \tau_1 + 2n \tau_2 $ & Yes if $(n \ne 1)$  \\ \hline
   $(n,3n)$ chiral & $ \langle \frac{5n}{3} \rangle +n \tau_1 + 3n \tau_2 $ & Yes if $(n \ne 1)$ \\ \hline
   $(n,4n)$ chiral & $ 2n + n \tau_1 + 4n \tau_2 $ & No \\ \hline
  \end{tabular}
 \end{center}
 \label{table:list}
\end{table}
The left index in Table~\ref{table:list} indicates the chirality, where
$(n,0)$ is `zigzag', $(n,n)$ is `armchair' and the others are `chiral'
type~\cite{SDD}.
The low energy bands are classified as orbiting states or non-orbiting
states, depending on $\mu^0_\perp$ in the center and right indices. 
It is noted that when $d =1$, as for a $(7,4)$ chiral nanotube,
$\mu^0_\perp$ is always congruent to zero, because 
$6 + 7 \tau_1 + 4 \tau_2 = 0$ holds for $(\tau_1,\tau_2) = (6,-12)$.

Proceeding now to the lattice structure along the axis, we examine the
wave vector of the electron located nearest the Fermi points in the
low energy bands. 
After calculating, we obtain the condition for choosing the state.
We label the wave vector along the axis of that state by
$\mu_\parallel^0$.
This wave vector satisfies the relation
\begin{equation}
 \left\langle \mu_\parallel^0 \frac{d_R}{d_T} \right\rangle = m.
\end{equation}
Depending on $\mu_\parallel^0 d_R/d_T$, we divide the axial structures
into two classes as follows:
\begin{align}
  \begin{cases}
   \displaystyle \mu_\parallel^0 \frac{d_R}{d_T} = m, \ \  & \text{(m-class)} \\
   \displaystyle \mu_\parallel^0 \frac{d_R}{d_T} \ne m. \ \ & \text{(s-class)}
  \end{cases}
\end{align}
Now, we consider $d_R$ for the two cases $d_R=d$ and
$d_R=3d$~\cite{SDD}.
When $d_R = d$, because $m/d$ is always an integer, we can choose an
integer $\mu_\parallel^0 = d_T (m/d)$ that satisfies the condition for
the m-class.
Also, when $d_R = 3d$, if $d_T$ is a multiple of 3, $\mu_\parallel^0$ is
again in the m-class.
For all other cases $\mu_\parallel^0$ belongs to the s-class.
To summarize our classification, we have
\begin{align}
 & d_R = d, \ \ \text{(m-class)} \\
 & d_R = 3d \ \begin{cases}
	       d_T = 3a, \ \ & \text{(m-class)} \\
	       d_T = 3a + i \ (i = 1,2), & \text{(s-class)}
	      \end{cases}
\end{align}
where $a$ is an integer.
For the m-class, the persistent current is the standard one. 
That is, the persistent current does not differ from the standard
sawtooth curve with a period equal to the flux quantum. 
However, for the s-class, the electron near one of the Fermi points
reaches the Fermi level when the $\Phi_0/3$ flux is turned on, and the
electron near another Fermi point reaches the Fermi level when the
$-\Phi_0/3$ flux pierces the torus.
Therefore, the resultant current is given by a superposition of two
sawtooth curves whose origins (or zero amplitude positions) are shifted
in different directions by $\pm \Phi_0/3$. 
This phenomenon was observed numerically by Lin and Chuu~\cite{LC} and
is a consequence of the well-known fact that one third of zigzag
nanotubes are metallic and the other two-thirds are semiconducting.

\section{Kinematics of a twisted torus}\label{sec:twisted_torus}

\begin{figure}[htbp]
 \begin{center}
  \psfrag{a}{(a)}
  \psfrag{b}{(b)}
  \psfrag{C_h}{$C_h$}
  \psfrag{T_w}{$T_w$}
  \psfrag{A_C}{$A \cdot C_h$}
  \psfrag{A_Tw}{$A \cdot T_w$}
  \psfrag{A}{$A$}
  \psfrag{u}{$u$}
  \psfrag{d}{$d$}
  \includegraphics[scale=0.5]{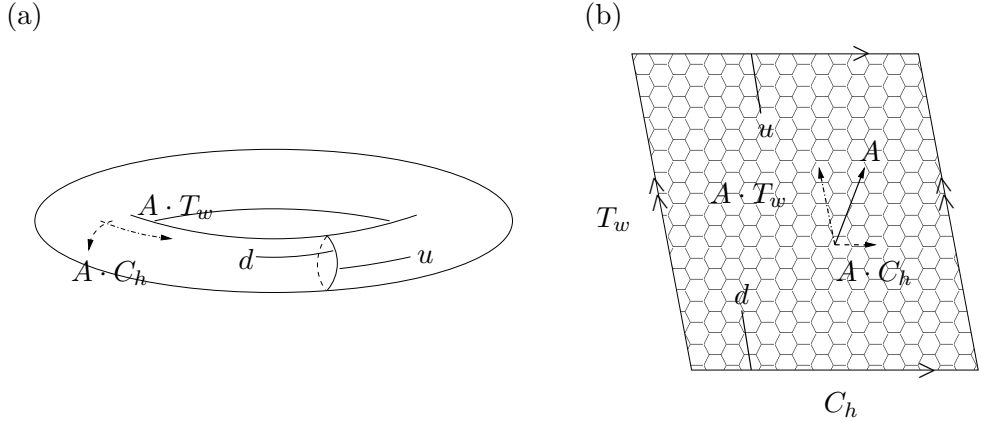}
 \end{center}
 \caption{Schematic diagram of a twisted torus (a) and its net (b) with
 an external gauge field. It is convenient to consider a parallelogram
 as the net of a twisted torus. Note that the two lines extending
 upward from `$u$' and downward from `$d$' are not joined, in contrast
 to the untwisted case.}
 \label{fig:twisted-torus}
\end{figure}

In the previous sections we examined the kinematics and persistent
currents in untwisted nanotori. 
Here, the twisted torus is investigated.
We define the translational vector for a twisted torus as $T_w$, which 
satisfies $C_h \cdot T_w \ne 0$.
All lattice structures except the untwisted torus belong to the twisted
nanotorus category (see Fig.~\ref{fig:twisted-torus}).
Among the various lattice structures of the twisted torus, we first
examine a type that can be obtained from an untwisted nanotorus. 
For this type, we can choose an untwisted torus having a translational 
vector that satisfies
\begin{equation}
  T_w - T \parallel C_h, 
  \label{eq:trans-twist}
\end{equation}
where $T$ is the translational vector of the corresponding untwisted 
nanotorus.
We call this an ``A-type'' twisted nanotorus.
We refer to another type of twisted nanotorus as ``B-type'', for
which we can also choose an untwisted nanotorus. 
It is defined in such a way that $|T_w-T|$ is of minimal length.
As an example of a B-type twisted nanotorus, we consider the zigzag
chiral vector $(n,0)$ and the translation vector $(d_t, -2d_t+1)$, where
$d_t$ is an integer.
The nanotorus is twisted because $C_h \cdot T_w = n T_1 \cdot T_2 \ne 0$.
Defining the corresponding untwisted torus for this twisted torus as $T =
d_t T_1 -2d_t T_2$, we then have $T_w - T = T_2$, which is not parallel
to $C_h (=nT_1)$. 
We limit ourselves to a study of A-type nanotori in this section.
We examine B-type nanotori in a subsequent section.
%%%%%%%%%%%%%%%%%%%%%%%%%%%%%
%%%%%%%%%%%%%%%%%%%%%%%%%%%%%

We choose the chiral and translational vectors of an A-type twisted
nanotorus as $C_h = n T_1 + m T_2$ and $T_w = p T_1 + q T_2$,
respectively, for the corresponding untwisted torus 
$T = \bar{p} T_1 + \bar{q} T_2$.
From Eq.~(\ref{eq:trans-twist}), we can relate $(p,q)$ to
$(\bar{p},\bar{q})$ using $(n,m)$ and an integer $t$ as  
\begin{equation}
 p = \bar{p} + t \frac{n}{d}, \ \ 
  q = \bar{q} + t \frac{m}{d}.
  \label{eq:twisted-torus}
\end{equation}
The integer $t$ determines the amount of twist, and the case of
vanishing $t$ corresponds to an untwisted torus.

To study the Hilbert space of conducting electrons in a twisted
nanotorus, we decompose the wave vector $k$ as 
$\mu_1 k_1 + \mu_2 k_2$,
where the unit wave vectors $k_1$ and $k_2$ are defined as
\begin{equation}
 C_h \cdot k_1 = 2\pi,\ C_h \cdot k_2 = 0,\  
  T_w \cdot k_1 = 0, \ T_w \cdot k_2 = 2\pi.
  \label{eq:k_2}
\end{equation}
The unit vectors are rewritten in terms of the reciprocal lattice
vectors as
\begin{equation}
 k_1 = \frac{1}{N_c} (-q K_1 + p K_2), \ 
  k_2 = \frac{1}{N_c} (m K_1 - n K_2).
  \label{eq:k_2_twist}
\end{equation}
Note that for an A-type torus, $N_c$ is still an even number, and it is
equal to that of the corresponding untwisted nanotorus
($N_c = mp-nq = m\bar{p} - n \bar{q}$).
As in the case of the untwisted nanotorus, $\mu_1$ and $\mu_2$ have the
following congruent degree of freedom:
$\delta \mu_1 = n \tau_1 + m \tau_2$ and $\delta \mu_2 = p \tau_1 + q
\tau_2$ where $\tau_1$ and $\tau_2$ are integers.

\section{Persistent currents in a twisted torus}\label{sec:PCinTT}

In this section, we examine persistent currents in twisted nanotori. 
A primary conclusion is that persistent currents depend on the degree of
twist.
This dependence is determined by the following two factors:
(1) whether or not there is a periodic lattice structure around the
axis, that is, whether $d \ge 2$ or $d = 1$, and 
(2) whether or not electrons near the Fermi level orbit about the axis.
This orbital degree of freedom is a new characteristic that
conventional one-dimensional material (such as a chain of atoms) does
not possess.

We start by specifying the low energy bands in twisted nanotori.
Even with twisted nanotori, the index of the low energy bands is
determined by the same conditions as for untwisted torus
(Eq.~(\ref{eq:rotation})).  
We choose the energy band index $\mu_1^0$ corresponding to a
non-orbiting state:
\begin{equation}
 \exists (\tau_1,\tau_2) \in Z \ \cdot \ \mu_1^0 =
  \left\langle \frac{2n + m}{3} \right\rangle + n \tau_1 + m \tau_2= 0.
\end{equation}
There are two low energy bands, for which the wave vectors around the
axis are given by $\pm \mu_1^0 k_1$. 
To examine the persistent currents, we consider the energy dispersion
relationship for the low energy bands,
\begin{equation}
 E_\pm = - V_\pi \left|\sum_{a=1,2,3} z_{a \pm}\right|
  = - V_\pi 
  \sqrt{ 3 + 2 \Re 
  \left[
   z_{1\pm}^* z_{2\pm} + z_{1\pm}^* z_{3\pm} + z_{2\pm}^* z_{3\pm}
  \right]
  },
\end{equation}
where we have defined
\begin{equation}
 z_{a \pm} \equiv \exp \left( i( \pm \mu_1^0 k_1 + \mu_2 k_2 - eA) \cdot u_a \right).
  \label{eq:block}
\end{equation}
It is not necessary to estimate the numerical value of $E_\pm$ to
understand the behavior for twisted nanotori.
It is enough to understand the difference between persistent currents in
untwisted and twisted tori. 
For this purpose, it is convenient to rewrite the vector $u_a$ in terms
of the chiral and translational vectors as 
\begin{align}
 &
 u_1 = \frac{2}{3N_c} 
 \left[ \left(p+\frac{q}{2}\right) C_h - \left(n+\frac{m}{2}\right) 
 T_w \right], \nn \\
 & 
 u_2 = \frac{2}{3N_c} 
 \left[ -\left(\frac{p-q}{2}\right) C_h + \left(\frac{n-m}{2}\right)
 T_w \right], \nn \\
 & 
 u_3 = \frac{2}{3N_c} 
 \left[ -\left(\frac{p}{2} + q \right)C_h + \left(\frac{n}{2} + m \right) 
 T_w \right]. \label{eq:u_2} 
\end{align}
In the absence of an external magnetic flux within the surface of the
nanotorus, we can set the gauge so that $A \cdot C_h = 0$.
Substituting Eq.~(\ref{eq:u_2}) into Eq.~(\ref{eq:block}) and using
Eq.~(\ref{eq:k_2}), we obtain 
\begin{align}
 &
 z_{1\pm}^* z_{2\pm} = 
 e^{-i(\pm 2\pi \mu_1^0)\frac{p}{N_c}}
 e^{i(2\pi \mu_2 -eA \cdot T_w) \frac{n}{N_c}}, \nn \\
 &
 z_{1\pm}^* z_{3\pm} = 
 e^{-i(\pm 2\pi \mu_1^0)\frac{p+q}{N_c}}
 e^{i(2\pi \mu_2 -eA \cdot T_w) \frac{n+m}{N_c}}, \nn \\
 &
 z_{2\pm}^* z_{3\pm} = 
 e^{-i(\pm 2\pi \mu_1^0)\frac{q}{N_c}}
 e^{i(2\pi \mu_2 -eA \cdot T_w) \frac{m}{N_c}}.
\end{align}
For an A-type twisted nanotorus, we can further rewrite the above
equations using Eq.~(\ref{eq:twisted-torus}) as
\begin{align}
 &
 z_{1\pm}^* z_{2\pm} = 
 e^{-i(\pm 2\pi \mu_1^0)\frac{\bar{p}}{N_c}}
 e^{i(2\pi \mu_2 -eA \cdot T_w \mp 2\pi \mu_1^0 \frac{t}{d}) 
 \frac{n}{N_c}}, \nn \\
 &
 z_{1\pm}^* z_{3\pm} = 
 e^{-i(\pm 2\pi \mu_1^0)\frac{\bar{p}+\bar{q}}{N_c}}
 e^{i(2\pi \mu_2 -eA \cdot T_w \mp 2\pi \mu_1^0 \frac{t}{d}) 
 \frac{n+m}{N_c}}, \nn \\
 &
 z_{2\pm}^* z_{3\pm} =
 e^{-i(\pm 2\pi \mu_1^0)\frac{\bar{q}}{N_c}}
 e^{i(2\pi \mu_2 -eA \cdot T_w \mp 2\pi \mu_1^0 \frac{t}{d})
 \frac{m}{N_c}}.
\end{align}
Comparing this with the wave vector in an untwisted nanotorus (the $t=0$ 
case), we observe that the twist produces a shift of the wave vector
along the axis. 
This effect can be regarded as a kind of gauge field induced by the
twist,~\footnote{A similar gauge field (commonly known as {\it
geometry-induced gauge}) was obtained by S. Takagi and
T. Tanzawa.~\cite{TT}}
assuming that the sign of the electron {\it charge} depends
on the orbital motion about the axis.
The twist induces a gauge field $A^{\rm twist}$.
Electrons in the low energy bands having charge ($e_\pm$) couple to
$A^{\rm twist}$, and therefore experience the effect of the total gauge
field as $A \to A + A^{\rm twist}$, where  
\begin{equation}
 e A \cdot T_w = 2\pi \frac{\Phi}{\Phi_0}, \ \ 
  e_{\pm} A^{\rm twist} \cdot T_w \equiv \pm 2\pi \mu_1^0 \frac{t}{d}.
\end{equation}
Here, $\Phi$ is the Aharonov-Bohm flux piercing the twisted torus.

\begin{figure}[htbp]
 \begin{center}
  \psfrag{I}{$I_{\rm pc}$}
  \psfrag{F}{$\Phi/\Phi_0$}
  \psfrag{I+}{$+\mu_1^0$ mode}
  \psfrag{I-}{$-\mu_1^0$ mode}
  \psfrag{a}{(a)}
  \psfrag{b}{(b)}
  \psfrag{c}{Add a Twist}
  \psfrag{d}{(c)}
  \psfrag{Total}{Total Current}
  \psfrag{0}{\tiny 0}
  \psfrag{1}{\tiny 1}
  \includegraphics[scale=0.3]{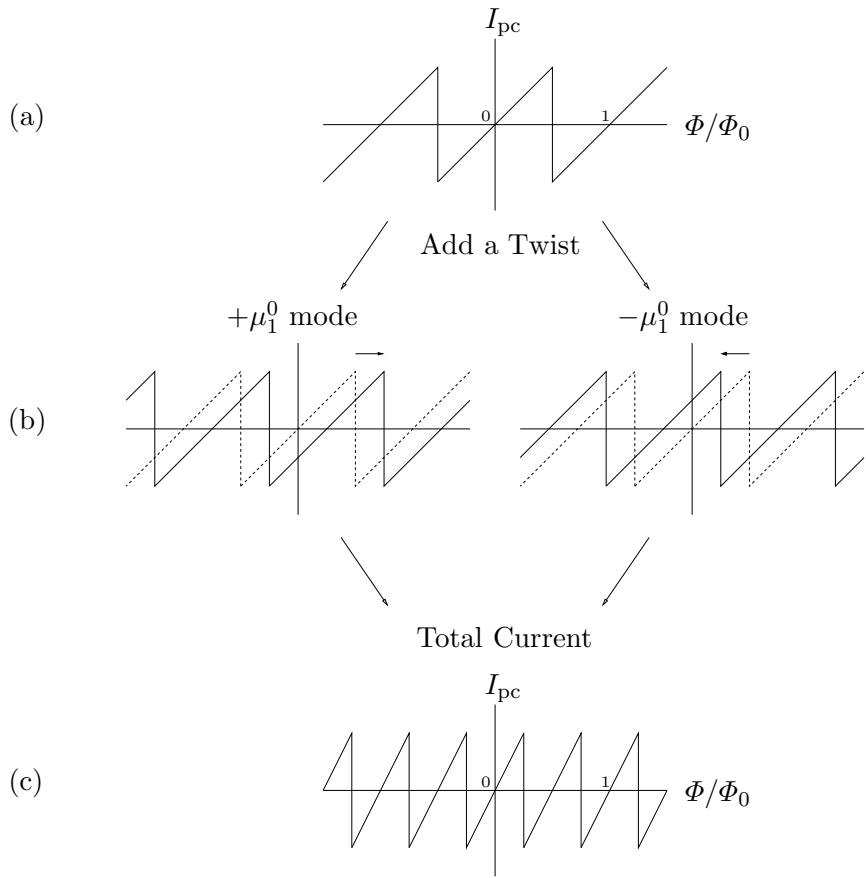}
 \end{center}
 \caption{(a) Sawtooth curve of a persistent current ($I_{\rm pc}$) in
 an untwisted (m-class) torus as a function of the magnetic flux. 
 The period of the curve is given by the flux quantum $\Phi_0$.  
 (b) The twist shifts the curves in the positive and negative
 directions, depending on the directions of the electrons' motion around
 the axis, i.e. clockwise or counterclockwise.
 The total persistent current is given by the sum of two currents. As an
 example, we depict the case of Eq.~(\ref{eq:half-period}) in (c).}  
 \label{fig:twist-pc}
\end{figure}

Let us verify the above result for an A-type twisted zigzag nanotorus
for which the chiral and translational vectors are defined by $(n,0)$
and $(p,q)$, where $p = \bar{p} + n_t$ and $q = \bar{q}$.
The constant $n_t$ represents the number of hexagons twisted at the
junction [see Fig.~\ref{fig:twisted-torus}(a)].
In this case, we have
\begin{equation}
 e_\pm A^{\rm twist} \cdot T_w = \pm 2\pi \mu_1^0 \frac{n_t}{n}.
\end{equation}
Hence, for the $(9,0)$ chiral vector, because $\mu_1^0 = 6$, we obtain a
$\pm \frac{4\pi}{3} n_t$ shift in the persistent current. 
We can imagine that the $\frac{2}{3}n_t \Phi_0$ magnetic flux penetrates
the ring on the assumption that signs of the charges of the electrons in
the low energy bands depend on the orbital motion about the axis, with a
positive sign for the clockwise case and a negative sign for the
counterclockwise case or vis versa (see Fig.~\ref{fig:twist-pc}).
Contrastingly, for an armchair chiral structure, the low energy band
is always classified as a non-orbiting state, i.e., $\mu_1^0$ can be
regarded as zero (see Table~\ref{table:list}).
As a result, the twist does not affect the persistent current. 
In other words, electrons near the Fermi level do not have the charges 
couple to the twist.

%%%%%%%%%%%%%%%%%%%%%%%%%%%%%
%%%%%%%%%%%%%%%%%%%%%%%%%%%%%

We now consider some possible consequences of the twist-induced gauge
field. 
First, the period of the persistent current becomes one half the flux
quantum for some nanotori.
Conventional one-dimensional material exhibits a period equal to the flux
quantum, except in the case of a superconducting state, in which a
sawtooth curve with a period equal to one half the flux quantum is
expected.  
Due to its twisted lattice structure, a carbon nanotorus can exhibit a
persistent current, with a period one half the flux quantum, even though 
the electrons do not form a superconducting state.
The condition for a period of one half is given by
\begin{equation}
 e_\pm A^{\rm twist} \cdot T_w = \pm \frac{\pi}{2} + 2\pi j
  \leftrightarrow \mu_1^0 \frac{t}{d} = \pm \frac{1}{4} + j,
  \label{eq:half-period}
\end{equation}
where $j$ is an integer.
It should be mentioned that the phenomena of the half period may also be
realized in an untwisted torus when the two low energy bands,
respectively, have even and odd (or odd and even) numbers of
electrons.~\footnote{The one-half periodicity might be realized even in
a conventional one-dimensional material if we assume even-odd asymmetry
for spin up-down conducting electrons.
Note also that the one-half period phenomena were experimentally
observed in systems consisting of a large number of loops and are
believed to be due to their ensemble average~\cite{Webb}.}
However, in the case of the twisted nanotorus, the additional surprising
phenomenon of a vanishing external magnetic field is expected.
Let us consider an m-class untwisted torus and suppose that the number of
electrons is even in the $+\mu_1^0$ energy band and odd in the $-\mu_1^0$
energy band.
The zero amplitude positions of the persistent currents are different,
or the phases differ by $\pi$, so that we have one-half periodicity.
Then, by twisting the untwisted nanotorus, the persistent currents of
both modes shift in different directions, and the amplitude of the total 
current has a chance to attain a finite value.
In this case, the persistent current can flow in the twisted torus
with no external magnetic field.

We now examine the case of B-type twisted nanotori.
We write the translation vector for a B-type twisted torus as $(p,q)$
and the corresponding untwisted torus as $(\bar{p},\bar{q})$ and express
their relation as $p = \bar{p} + \delta p$ and $q = \bar{q} + \delta q$.
After calculating, we obtain
\begin{align}
 & 
 z_{1\pm}^* z_{2\pm} = 
 e^{-i\frac{2\pi}{N_c} 
 \left( \pm \mu_1^0 p - \frac{eA \cdot C_h}{2\pi}\bar{p}
 - \mu_2 n + \frac{e A\cdot T_w}{2 \pi} n \right)}, \nn
 \\
 &
 z_{1\pm}^* z_{3\pm} = 
 e^{-i\frac{2\pi}{N_c} 
 \left( \pm \mu_1^0 (p+q)- \frac{eA \cdot C_h}{2\pi} (\bar{p} + \bar{q})
 - \mu_2 (n+m)
 + \frac{e A \cdot T_w}{2 \pi} (n+m) \right)}, \nn \\
 &
 z_{2\pm}^* z_{3\pm} =
 e^{-i\frac{2\pi}{N_c}
 \left( \pm \mu_1^0 q - \frac{eA \cdot C_h}{2\pi} \bar{q}
 - \mu_2 m + \frac{e A \cdot T_w}{2 \pi} m \right)}.
\end{align}
Proceeding further, we set 
$\delta p T_1 + \delta q T_2 = \alpha C_h + \beta T$,
where $\alpha$ and $\beta$ are fractions.
Note that, $\beta=0$ ($\beta \ne 0$) for an A-type (B-type)
twisted nanotorus.
Using this definition, we obtain
$\delta p = \alpha n + \beta \bar{p}$ and 
$\delta q = \alpha m + \beta \bar{q}$.
It is clear that the effect of the $\beta$ term can be absorbed into the
effective gauge field as before, but this time the inner product of
$A^{\rm twist}$ and the chiral vector does not vanish:
\begin{equation}
 e_{\pm} A^{\rm twist} \cdot T_w \equiv \pm 2\pi \mu_1^0 \alpha, \ 
  e_{\pm} A^{\rm twist} \cdot C_h \equiv \pm 2\pi \mu_1^0 \beta.
  \label{eq:A_twist_radial}
\end{equation}
Note that $|\beta| \ll |\alpha|$ holds for lattice structures
satisfying $|T_w -T| \le |C_h|$ and $|T| \gg |C_h|$. 
The effect of the $\beta$ term on the electrical properties will become 
important when we consider a small $|T| \sim {\cal O}(|C_h|)$
nanotorus.
This result indicates that a B-type twisted nanotorus may have a tiny
energy gap even if its corresponding untwisted torus is classified as a 
metallic system and m-class.

We now present a simple derivation of the final result of
Eq.~(\ref{eq:A_twist_radial}). 
For both twisted and untwisted tori, we have defined the unit wave
vectors as given in Eqs.~(\ref{eq:k_parallel_normal}) and
(\ref{eq:k_2_twist}).
They are related by
\begin{equation}
 k_1 = \frac{\bar{N_c}}{N_c} k_\perp 
  + \frac{1}{N_c}\left( - \delta q K_1 + \delta p K_2 \right),\  
  k_2 = \frac{\bar{N_c}}{N_c} k_\parallel,
\end{equation}
where $\bar{N}_c \equiv m\bar{p}-n\bar{q}$.
It is noted that $N_c = \bar{N}_c$ holds for A-type twisted nanotori.
However, no such relation is believed to exist for the B-type torus.
The key physical quantity appears on the right-hand side of the first
equation, where   
\begin{align}
 \frac{1}{N_c}\left( - \delta q K_1 + \delta p K_2 \right) 
 &= \frac{1}{N_c} 
 \left[ 
 \left( -n \delta q + m \delta p \right) k_\perp
 +\left( -\bar{p} \delta q + \bar{q} \delta p \right) k_\parallel
 \right] \nn \\
 &=
 \frac{\bar{N}_c}{N_c} \left[ \beta k_\perp - \alpha k_\parallel \right].
\end{align}
The above equation indicates that the wave vectors around and along the
axis change according to the twist, represented by $\beta$ and $\alpha$.
These terms can be thought of as an effective gauge field due to the
gauge coupling, and this is why persistent currents exhibit a
distinctive shape.

\section{Summary and discussion}\label{sec:discussion}

We have studied the kinematics of $\pi$-electrons in nanotori and have
shown that their persistent currents depend on the lattice structure. 
In particular, we have clarified the effects of twist on the persistent
currents and have revealed two consequences: the flux period can be one
half of the single-electron flux quantum, and a non-vanishing current
can flow without an external magnetic field.
We have shown that in order to observe the effect of twist on persistent
currents, the following geometrical conditions have to be satisfied:
(1) there must be a periodic lattice structure around the axis, that is,
$d ={\rm gcd}(n,m)$ is not unity; and
(2) the electrons in the low energy bands must be orbiting about the
axis.

In this paper, we did not consider the effect of Coulomb interactions
among conducting electrons on persistent currents, because this effect
is not well understood even for standard one-dimensional Aharonov-Bohm
rings.
However, there are several indications.
For instance, experimentally,
Chandrasekhar et al.~\cite{Webb} measured the magnetization of single,
isolated Au loops and observed that the amplitude of the persistent
currents is close to the value predicted by theories of non-interaction.
Theoretically, it is shown that for a one-dimensional model derived from 
the tight-binding Hamiltonian (Eq.~(\ref{eq:hamiltonian})), the
long-range part of the Coulomb interaction is ineffective, indicating
that persistent currents {\it persist} even in the presence of external
charges (Sasaki in Ref.\citen{LRR}).
Although our argument is based mainly on kinematics, it is possible that
interactions might invalidate our results.

In addition to Coulomb interactions, the Hamiltonian is expected to be
modified by several factors.
The surface curvature and bending of the tube is known to affect the
location of the Fermi points~\cite{KM}.
Furthermore, the Euler theorem for polyhedra permits pentagon-heptagon
pairs in nanotori, and a pentagon (or a heptagon) can mix the wave
functions at two Fermi points~\cite{GGV}.
Therefore, the persistent current is thought to be affected by their
presence.
This warrants future work concerning the effects of dynamical details
(surface curvature and so on) on the persistent current. 
However, note that the essential physics presented in this paper can
be applied to persistent currents in tubule structures based not only on
carbon but on other materials as well~\cite{SKS}.

\section{Conclusion}\label{sec:conclusion}

The kinematics of $\pi$-electrons in carbon nanotori have been clarified
and used to examine persistent currents.
We have shown that persistent currents in twisted nanotori exhibit a
characteristic different from that of conventional one-dimensional
materials due to the fact that conducting electrons near the Fermi level
orbit about the axis. 
The results clearly show that the lattice structure itself allows the
appearance of new phenomena in the persistent currents.

We can now answer the question of how nanotubes differ from conventional
one-dimensional materials, such as a chain of atoms?, posed in
\S~\ref{sec:introduction}.
The difference is that conducting electrons near the Fermi level have an
orbital degree of freedom about the nanotube tubule axis, and the
nanotube characteristics determine the persistent currents in the
nanotori.

\section*{Acknowledgements}
K. S. is grateful to Dr. Y. Sumino for his continuous encouragement
and outstanding instruction during the preparation of this paper.
He also wishes to thank Prof. Z. F. Ezawa, Prof. K. Hikasa,
Prof. N. Toyoda, and Prof. Y. Kuramoto.
He would like to thank the members of the High Energy Theory
Group at Tohoku University, including Prof. S. Watamura, Prof. T. Moroi,
Dr. M. Hotta, and Dr. H. Ishikawa.
This work is supported by a fellowship of the 21st Century COE Program
of the International Center of Research and Education for Materials of
Tohoku University.

%%%%%%%%%%%%%%%%%%%%%%%%

\end{document}